\documentclass[fleqn,twoside]{article}
\usepackage{espcrc2}
\usepackage[dvips]{graphicx}
\usepackage{latexsym}

\author{Matthew A. Nobes\address[sfu]{Physics Department, Simon Fraser University, Burnaby, BC, Canada}, 
Howard D. Trottier\addressmark}

\title{Progress in automated perturbation theory for heavy quark physics}
\begin{document}

\begin{abstract}
We review our perturbative techniques for improved heavy quark actions.  A new
procedure for computing improvement coefficients is suggested, where the continuum limit of a lattice-regularized
theory provides the matching conditions.We also use a gauge-invariant infrared regulator that is well
suited to higher-order calculations in lattice gauge theory.
We report on preliminary tests of our method, as well as a possible way 
to reduce systematic errors in these calculations.
\vspace{1pc}
\end{abstract}

\maketitle

\section{Introduction}

Improved actions are widely used in heavy quark physics.  Examples
include NRQCD \cite{lepage1} which is very successful in describing $b\bar{b}$ systems, and the Fermilab 
action \cite{kkm} which has been applied to both
$b$ and $c$ quark systems.  With the MILC results for unquenched light quarks, precision 
determinations of important physical quantities (i.{}e.{} $f_{B}$) are possible with these 
heavy quark actions \cite{hpqcdpre}.

One of the dominant systematic errors that remains with improved actions is the determination of the
improvement coefficients. The hyperfine splittings of the J/$\psi$ system illustrates this clearly.
While there have been good results from heavy quark simulations, these hyperfine 
splittings remain significantly underestimated with tree-level improved actions.
This underestimate is seen with both the Fermilab and NRQCD approaches to heavy quark physics.
Unquenching (\cite{roman},\cite{paul02}) does not seem to resolve it (the Fermilab
determination is still $\approx$ 25\% low).  We expect that
this underestimate should be greatly reduced by perturbative matching of  
the $\Sigma \cdot \vec{B}$ operator and the use of a highly improved action \cite{bugra}.

In order to determine this action coefficient ($c_{B}$) we adopt the simplest matching strategy 
possible. We will compute the amplitude for quark scattering off of a background
chromo-magnetic field in lattice perturbation theory, and, 
to the same order in perturbation theory compute the continuum result,
using the same IR regulator.  The matching is simply a matter of subtracting one from the 
other, and tuning $c_{B}$ to insure that the difference vanishes to whatever order is desired

In principle, this procedure should work to whatever order in perturbation theory 
we wish.  In this report we will discuss the matching to one-loop order, however
for true high precision results the matching should be done to two loops.
This is because at typical lattice spacings $\alpha_{V}\left(1/a\right) \approx 0.2$.
Our ultimate goal is high precision results, so we want to organize our calculation
in a way which makes going to two loop order as easy as possible.

We have discussed our approach to lattice perturbation theory at length in
prior LATTICE presentations (\cite{lat01},\cite{lat02},\cite{chief}).  The core of our technique 
is to generate Feynman rules automatically using the L\"uscher - Weisz algorithm \cite{lw}.
The primary advantage of this method is that we can change the form of the action easily, despite 
the complexity of the rules.

Many of the individual diagrams we are calculating are infrared divergent, to regulate this we use
twisted boundary conditions.  This has the effect of turning integrals over the three momentum into 
restricted sums, as in the following example:
{\small
$$
\int dk_{0} d\vec{k} \frac{f(k)}{\hat{k}^{2}} \to \int dk_{0} \sum_{\vec{k}} \frac{\chi_{k}f(k)}{\hat{k}^{2}}.
$$
}
Here $\chi_{k}$ is a veto function which excludes many lattice momenta, including the IR zero modes.

\begin{figure}[!h]
\begin{center}
\includegraphics[totalheight=7cm,width=6cm]{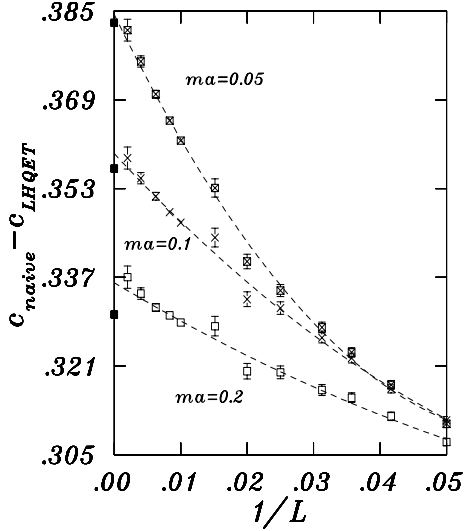}
\caption{Lattice to Lattice Matching Test}
\label{l2lmatch}
\end{center}
\end{figure}

\section{Lattice-to-Lattice Matching}

By design, the lattice and continuum theories should be identical in the infrared. 
The lattice matching computations that have been done to date have typically
used a gluon mass to regulate the IR in both the lattice and continuum theories.
This method could work for our calculation, however, it may pose significant problems with 
gauge invariance at higher order.

One solution is to use twisted boundary conditions, which provide a gauge invariant IR regulator. 
However this regulator is somewhat difficult to combine with dimensional regularization for 
continuum calculations.
This can be avoided by rethinking the continuum UV regulator 
that we use.  We are calculating matching coefficients, so there is no particular reason
to use dimensional regularization.  In fact, we can use any UV regulator we want.  

Following a suggestion of Peter Lepage we have simply used a lattice cutoff
to regulate the continuum theory.  This allows us to do the ``continuum'' side of the computation 
using the same lattice techniques as we have developed for the heavy quark actions.
This procedure is straightforward, we can pick a simple lattice theory (Wilson glue +
naive quarks) and run the lattice spacing down.  As long as the spacing is made small 
enough, this theory will be very close to the continuum theory.  Then we can subtract
off our heavy quark results, and get the matching coefficients.
This lattice-to-lattice technique should work for matching                                                                
calculations within QCD itself, such as the perturbative coefficients for
operators in the lattice action. This method does not apply for example
to matrix elements of operators in the effective Hamiltonian for the weak interactions.

As we discuss below, for one loop matching computations this naive procedure works.
To two (and higher) loops the ``continuum'' and lattice theories will both have to be 
renormalized to give 
the same physical values of the various parameters, however this should be 
a straightforward application of renormalized perturbation theory.

\begin{figure}[!t]
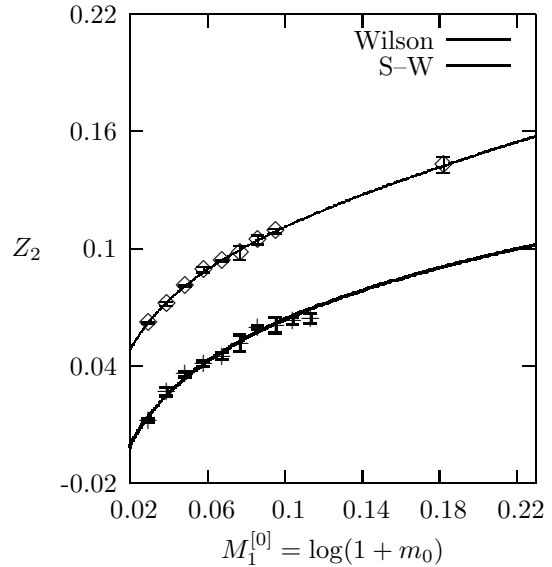

\begin{center}
\include{z2light}
\caption{Wavefunction renormalization for light Wilson and SW quarks}
\label{z2}
\end{center}
\end{figure}

\section{Testing}

To test this procedure we do the computation of $c_{\Sigma\cdot B}$ in the static 
quark limit. This calculation has previously been performed, using a gluon mass as
the IR regulator, and matching to the dimensionally regularized continuum
theory, by Flynn and Hill \cite{fh}. We compute the 
amplitude for quark scattering off of a background field
for both the heavy quark and ``continuum'' theories.

In the ``continuum'' case, we're using units in which $a=1$ so 
we want to run all other scales in the problem down to zero. 
For matching, we use naive quarks, so the differences from the continuum limit
go like $\mathcal{O}(m^{2})$.

Figure \ref{l2lmatch} shows the results of this lattice to lattice matching for various naive quark
masses.  The black squares 
are the results of \cite{fh}.  Clearly, in the continuum limit, $L\to\infty$ and $m \to 0$, 
we reproduce 
the Flynn-Hill result.  Figure \ref{l2lmatch} shows that our 
matching procedure works, and that the errors will be small enough
to reliably extract the dependence on the ultraviolet cutoff.

In addition to this test, we have throughly 
tested our implementation of the Fermilab fermion action.  To do this we have duplicated
all of the major results in \cite{mkk} for the one loop rest mass and wavefunction 
renormalization. Figure \ref{z2} shows the (one loop)
wavefunction renormalizations of Wilson and clover quarks, at fixed lattice size.
The logarithmic dependence on $M_{0}$ agrees with \cite{mkk}.

\section{Periodic Boundary Conditions}

We have investigated other means for controlling infrared
divergences. One possibility is to integrate the difference between the
lattice and the "continuum" expressions for a matching coefficient. 
This difference is infrared finite, and hence could be computed without any 
infrared regulator.

To test this method, we have calculated the difference in fermion wave function renormalizations
$Z_{2}^{Wilson}-Z_{2}^{SW}$ over a range of fermion masses, without any infrared regulator.
In the $m_{0}\to 0$
limit we obtain $0.228(1)$ which is in reasonable agreement 
with $0.023083(7)$ which is the result presented
in \cite{mkk}.  We are
currently investigating this technique for determining $c_B$.

\section{Conclusions}

Doing high precision QCD with improved heavy quark actions requires two loop
determinations of the improvement coefficients.  In particular
the coefficient of the $\sigma \cdot B$ is needed to resolve the underestimate of the
hyperfine splitting we have discussed. 

To get to two loops it is important to organize the calculation in
as efficient a manner as possible.  To this end we have introduced a procedure
to match two lattice theories onto each other, where the continuum limit of 
one lattice regularized theory provides the matching
criteria for the improvement of the other. Additionally we have used
twisted boundary conditions to provide a gauge invariant regulator. As a 
preliminary test of these techniques we have reproduced an
earlier result for the matching of $\sigma \cdot B$ in the lattice
HQET theory. We are currently applying these methods to improvement of 
the Fermilab and NRQCD actions.


\begin{thebibliography}{99}
\bibitem{lepage1} G.P. Lepage, \emph{et. al.} Phys.Rev. D46 (1992) 4052-4067.
\bibitem{kkm} A. El-Khadra, \emph{et. al.} Phys. Rev. D55 (1997) 3933-3957.
\bibitem{hpqcdpre} G.P. Lepage, \emph{et. al.} hep-lat/0304004
\bibitem{roman} R. Koniuk and C. Stewart. Nucl.Phys.Proc.Suppl. 94 (2001) 375-378.
\bibitem{paul02} M. Di Pierro , \emph{et. al.} Nucl.Phys.Proc.Suppl. 119 (2003) 586-591.
\bibitem{bugra} M. Oktay.  These proceedings.
\bibitem{lat01} M. Nobes, \emph{et. al.} Nucl.Phys.Proc.Suppl. 106 (2002) 838-840.
\bibitem{lat02} M. Nobes, H. Trottier. Nucl.Phys.Proc.Suppl. 119 (2003) 461-463.
\bibitem{chief} H. Trottier. These proceedings.
\bibitem{lw} M. L\"uscher and P. Weisz. Nucl. Phys. B266 (1986) 309-356.
\bibitem{fh} J. Flynn, B. Hill. Phys. Lett. B264 (1991) 173-177.
\bibitem{mkk} B. Mertens, \emph{et. al.} Phys. Rev. D58, 034505.
\end{thebibliography}
\end{document}